\documentclass[preprint,12pt]{elsarticle}

\usepackage[utf8]{inputenc} 
\usepackage[T1]{fontenc}  
\usepackage{amsmath,amssymb,amsfonts}%
\usepackage{amsthm}%
\usepackage{algorithm}%
\usepackage{algpseudocode}%
\usepackage{listings}%
\usepackage{hyperref}%
\usepackage{caption}%
\usepackage{placeins}%

\journal{Simulation Modelling Practice and Theory}

\makeatletter
\def\ps@pprintTitle{%
  \let\@oddhead\@empty
  \let\@evenhead\@empty
  \def\@oddfoot{\hfill}%
  \let\@evenfoot\@oddfoot}
\makeatother

\begin{document}

\begin{frontmatter}



\title{A Graph-Based Laser Path Solver Algorithm for Virtual Reality Laboratory Simulations}



\author[1,2]{Andreas M{\"u}ller}

\author[2]{Stefan M{\"u}ller}

\author[2]{Tobias Brixner}

\author[1]{Sebastian von Mammen}

\affiliation[1]{organization={Institut f{\"u}r Informatik, Universit{\"a}t W{\"u}rzburg},
            addressline={Am Hubland},
            city={W{\"u}rzburg},
            postcode={97074},
            state={Bavaria},
            country={Germany}}

\affiliation[2]{organization={Institut f{\"u}r Physikalische und Theoretische Chemie, Universit{\"a}t W{\"u}rzburg},
            addressline={Am Hubland},
            city={W{\"u}rzburg},
            postcode={97074},
            state={Bavaria},
            country={Germany}}

\begin{abstract}
femtoPro is an interactive virtual reality (VR) laser laboratory balancing the contrasting challenges of accuracy and computational efficiency in optics simulations.
It can simulate linear and nonlinear optical phenomena in real time, a task that pushes the boundaries of current consumer hardware.
This paper details the concept, implementation, and evaluation of a dynamic graph-based solution tailored to the specific requirements and challenges of the simulation.
Resource usage is optimized through a selective updating strategy that identifies and preserves laser paths unchanged between simulation frames, eliminating the need for unnecessary recalculations.
Benchmarking of real-world scenarios confirms that our approach delivers a smooth user experience, even on mobile VR platforms with limited computing power.
The methodologies, solutions and insights outlined in this paper may be applicable to other interactive, dynamic graph-based real-time simulations.
\end{abstract}


\end{frontmatter}



\section{Introduction}
\label{sec:introduction}

femtoPro is an interactive laser laboratory simulation, enabling the safe conduct of complex experiments in an immersive virtual reality environment as we introduced elsewhere \cite{femtoProMain2023}.
It features a realistic optical model that contains a mixture of geometrical and wave-optics characteristics to facilitate an accurate simulation in real time, ensuring a stable frame update rate of at least 90~frames per second (fps).
A detailed discussion on the physical realism and accuracy of the underlying optical models is provided in a separate publication \cite{femtoProSimulation2025}.
Current methods for simulating light-wave propagation and interaction with matter use a variety of different approximations and models for a solution of Maxwell's equations, with some also integrating quantum mechanical principles \cite{SIM2009,QUANTUM,LUMERICAL,Meep2010}.
Due to their significant computational demand, Maxwell solvers of high accuracy are typically used for designing optical experiments or simulating experimental data a-posteriori.
In contrast, in a VR application, real-time calculation is essential, setting distinct requirements for femtoPro.

Ray-tracing algorithms on GPU hardware have become highly efficient for simulating the behavior of light in real time, now even capable of handling computationally intensive phenomena like the dispersion of white light into color spectra and caustic effects \cite{Raytracing2021}.
Rather than tracing millions of rays in parallel, femtoPro simulates the propagation of one beam with a Gaussian intensity cross-section profile as it interacts with a sequence of optical elements (e.g., mirrors and lenses) in its pathway.
A section of that path leading from one point of interaction (e.g., the laser source) to the next (e.g., a mirror) is defined as a beam segment.
In order to implement femtosecond laser pulse properties, each beam segment contains complex-valued array structures that represent the 
frequency and time distributions of the electric field. 
This allows us to incorporate amplitude and phase information and handle interference phenomena.
Given the specific needs to simulate these detailed laser pulse properties, we opted for a CPU-based algorithm over ray tracing on the GPU to efficiently manage associated calculations.
While GPUs are optimized for highly parallel, uniform workloads, they are less suited for tasks involving tightly coupled data and dynamic structural changes. femtoPro's evolving beam graph and interdependent field calculations require flexible memory handling and sequential control, making CPUs better suited to this class of simulation.

The simulation is backed by a dynamic graph model that maps the formal network between beam segments of laser pulses propagating freely in space as edges, and optical elements (such as mirrors or lenses) that modify the laser pulse properties, as nodes.
The graph's topology evolves dynamically in real time, shaped by the underlying physical laws and interactions governing the system.
Research into dynamic graph models in general has gained significant traction only recently, with a notable surge in interest observed during the COVID-19 pandemic \cite{COVIDGRAPHS}.
The complexity of dynamic graph problems, potentially reaching NP-hardness in general cases, poses unique computational challenges and opens innovation avenues in multiple domains \cite{DynamicGraphHardness,ERLEBACH20211}.
However, the specific problem addressed in femtoPro does not involve NP-hard complexity, as our current implementation follows a structured and constrained traversal approach tailored to real-time optical simulation. Future extensions may explore traversal optimizations that resemble NP-hard problems such as the Traveling Salesman Problem, for example to minimize redundant computations in complex nonlinear configurations \cite{travelingSalesmanProblem}.
We use an interdisciplinary approach, combining graph theory, optical physics, and practical optimization techniques
to develop a solution tailored to the specific needs of an interactive real-time VR laser laboratory.
By sharing insights into its conceptual framework, implementation, and performance evaluation, femtoPro not only showcases a distinctive application of dynamic graphs in real-time settings, but also highlights both challenges and solutions related to performance within such systems, serving as a reference for researchers and developers seeking to apply similar approaches in their work.

Following this introduction, Section~\ref{sec:requirements} provides an overview of the requirements for the femtoPro laser simulation.
How these requirements are met is then detailed in Section~\ref{sec:implementation}, introducing the implementation of femtoPro's dynamic graph model
and examining the laser path solver algorithm, crucial for the graph to continuously and accurately represent the system's current state.
The performance of the simulation is analyzed in Section~\ref{sec:analysisMain}.
In Section~\ref{sec:analysisInGeneral}, equations for computational costs are derived, providing the basis for a complexity analysis as a first indicator for the system's scalability.
Section~\ref{sec:analysisOfExperiments} addresses how performance is affected in real-world scenarios, alongside a presentation and analysis of benchmark results of typical experimental setups.
Section~\ref{sec:analysis_practicalConsiderations} explores the practical implications of algorithmic performance on user experience.
Section~\ref{sec:conclusion} concludes this paper with final thoughts on our results and an outlook on planned features of femtoPro.

\section{Requirements}
\label{sec:requirements}

This section outlines the features and requirements of femtoPro \cite{femtoProArchitecture2023,femtoProMain2023} that form the basis for the underlying graph model and its implementation.

femtoPro dynamically calculates and renders the path of laser beams in real time as they traverse through free space and various optical elements.
Each beam has specific properties that determine its appearance and enable a detailed analysis.
For example, the frequency envelope captures the energy distribution at a given point in space across various frequencies of a specific sampling size. 
As another example, the time envelope represents the temporal characteristics of a pulsed laser beam where the electric field amplitude changes over time.
Although in the real world, the path of a laser beam through air is typically not visible, it can be rendered as a volumetric 3D shape for didactic reasons.
Both the beam path and cross sections are visualized with color and transparency values corresponding to the beam's spectral and radial intensity distributions.

Optical elements can be divided into two types: apertures and optical media.
Iris apertures have a central opening whose size can be adjusted to control the amount of light that passes through.
Optical media consist of a material with geometrical attributes defining its front and back surface (e.g., radius, curvature and thickness), as well as physical parameters (e.g., reflectivity, transmittance, and index of refraction). 
These media can modify the spatial, frequency, and time properties of incident pulses.
Reflectivity causes part of the incoming light to be redirected from the optical element's surface, forming a reflected beam.
Transmittance allows a portion of the light to pass through, creating a transmitted beam.
For simplicity, femtoPro currently treats only the reflection at the incident interface of the medium and ignores a potential second reflection at the exit interface. 
The refractive index determines the direction of a beam and can be chosen as frequency-dependent to take dispersion into account.
Interaction with curved surfaces causes a beam to converge or diverge. 
This allows for focusing or defocusing via curved mirrors or lenses.
In summary, all attributes collectively determine how an incident beam is transformed into one or more outgoing beams.

femtoPro supports various types of optical media:
mirrors (reflecting beams),
lenses (primarily focusing or defocusing transmitted beams, though a fraction of light may also be reflected),
filters (transmitting only specific wavelength intervals),
beam splitters (partially transmitting and reflecting beams), 
beam blockers,
and measurement devices (spectrometers and power meters).
While spectrometers analyze the spectral intensity distribution, power meters measure the integrated power of a beam incident on a detecting surface.

A fundamental requirement of femtoPro is that experimental setups are fully interactive and simulated in real time, enabling immediate observation of the impacts of any interaction with the simulation.
Similar to a real-world laboratory, there are various ways to manually adjust the position and/or orientation of an optical element via its opto-mechanical holder. 
Fine alignment is performed via adjustment screws which
allow for vertical and horizontal tilt as well as displacement of an optical element.

\begin{figure}[H]
{
\centering
\includegraphics[width=1\linewidth]{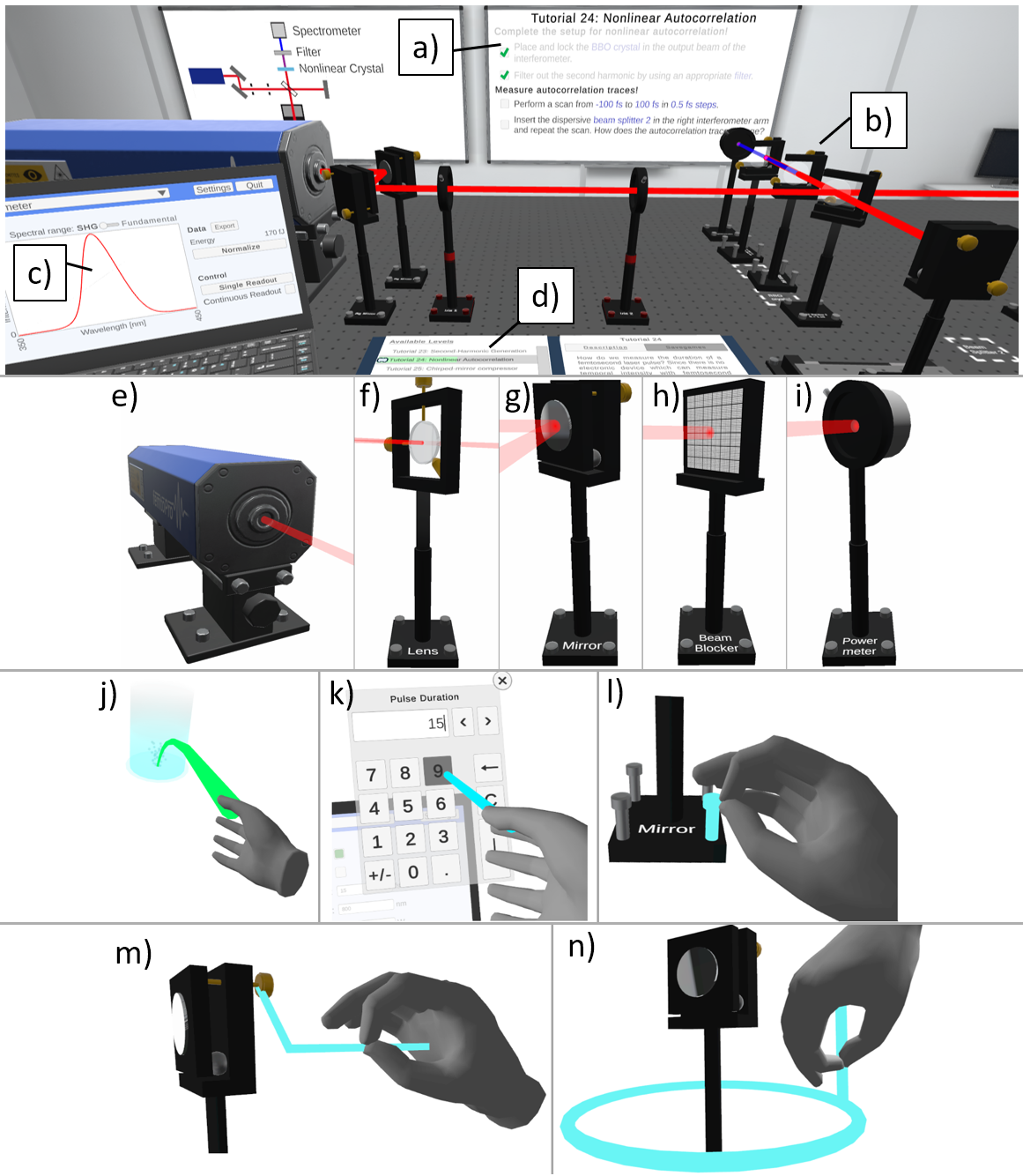}
\caption{Collection of screenshots showing femtoPro's major features and interaction elements: (a) didactical tasks on the whiteboard, (b) experimental setup, (c) virtual laptop for controlling and analyzing experiments, (d) level folder for loading, saving, and restarting experimental levels, (e)--(i) optical elements, such as (e) laser source, (f) lens, (g) mirror, (h) beam blocker, (i) power meter, (j)--(n) user interactions, such as (j) teleporting, (k) inputting a number, (l) locking a screw, (m) fine and (n) coarse adjustment of an optical element. For snippets (e)--(n), a white background was added to enhance visual clarity. Further details are provided elsewhere \cite{femtoProArchitecture2023}.
\label{fig:femtoPro_screenshots}}
}
\end{figure}

For better understanding, Figure~\ref{fig:femtoPro_screenshots} shows a collection of in-game screenshots illustrating femtoPro's core features in virtual reality as experienced by the user. The status of didactical tasks is presented on a whiteboard (Fig.~\ref{fig:femtoPro_screenshots}a). The freely-configurable experimental setup (Fig.~\ref{fig:femtoPro_screenshots}b) can be controlled and analyzed in various ways via a virtual laptop (Fig.~\ref{fig:femtoPro_screenshots}c). The available experimental levels are listed in a level folder, which allows loading, saving, and restarting any given level (Fig.~\ref{fig:femtoPro_screenshots}d). 
The virtual laptop, level folder, optical elements (e.g., Fig.~\ref{fig:femtoPro_screenshots}e--i), and measurement devices (e.g., Fig.~\ref{fig:femtoPro_screenshots}i) can be freely grabbed and placed on the experiment table.
Figure~\ref{fig:femtoPro_screenshots}j--n depicts further user interactions, such as navigating the lab environment via teleport (Fig.~\ref{fig:femtoPro_screenshots}j), inputting numerical values via a numpad interface (Fig.~\ref{fig:femtoPro_screenshots}k), locking an optical element onto the table via screws (Fig.~\ref{fig:femtoPro_screenshots}l) as well as fine (Fig.~\ref{fig:femtoPro_screenshots}m) and coarse (Fig.~\ref{fig:femtoPro_screenshots}n) alignment of an optical element. A more detailed explanation of these features can be found elsewhere \cite{femtoProArchitecture2023}.

As illustrated in Figure~\ref{fig:femtoPro_screenshots}, femtoPro offers a highly flexible, scalable, configurable, and freely interactable experimental environment. Users are not limited to predefined setups with fixed component positions; instead, they can assemble and precisely align any experiments with given optical elements, while still obtaining accurate results through our adaptable real-time physics simulation.
The physical accuracy of the underlying optical models has been discussed and validated in a separate publication \cite{femtoProSimulation2025}.

\begin{figure}[htb]
{
\centering
\includegraphics[width=0.5\linewidth]{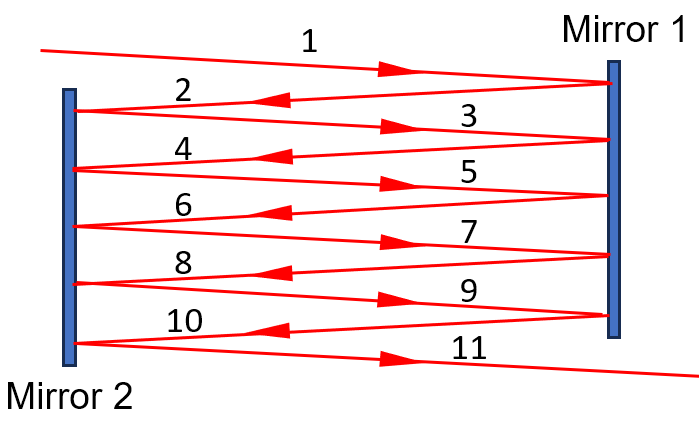}
\caption{Multiple-reflection scenario with red arrows depicting beam segments labeled in ascending order to illustrate the progression of the beam path.\label{fig:cavity_scheme}}
}
\end{figure}

femtoPro's goal to allow unrestricted exploration and experimentation through a freely-configurable experimental platform requires the capability of simulating any setup the users wish to construct, beyond just those predefined in literature.
An experimental setup might be aligned in such way that a beam repeatedly reflects between two or more optical elements in a recurring cycle.
In practice, particularly in optical cavities, the number of recurring reflections can approach infinity as the beam's energy is decaying exponentially with increasing number of reflections \cite{trager2012}.
Our model sets an arbitrary limit on the number of reflections that simplifies calculations but is not physically accurate.
Figure~\ref{fig:cavity_scheme} shows a finite bouncing scenario where each beam segment is annotated with an ascending number, indicating the unique and sequential path of laser light.

The optical media previously described adhere to linear optics principles, where beams spatially overlapping at the media's surface do not interact with each other, but instead preserve their individual properties.
A nonlinear crystal (NLC) is a transmissive optical medium that also supports second-order nonlinear optical responses such as sum-frequency or second-harmonic generation.
This means that two or more beams intersecting at the surface of an NLC may collaboratively create additional outgoing beams with frequencies different from those of the incident beams.
It is important to note that while an outgoing beam might appear collinear and thus seem to be directly influenced by an incident beam, our model considers each transmitted beam to arise independently from the combined effects within the medium \cite{femtoProMain2023}. 
This approach offers a more precise representation of nonlinear light-matter interactions than other models that might assume direct interactions between overlapping incident beams.

With an NLC, the parameters of each incident beam must be known and considered in relation to every other incident beam to calculate the paths and properties of any subsequent outgoing beams. 
The dependency of output beams on all combinations of input beams means that parallel computation, common in linear optics modeling, is less feasible. 
Thus, the capability of handling nonlinear optics significantly impacts how the simulation is modeled and implemented, increasing its computational load and complexity.

\begin{figure}[htb]
{
\centering
\includegraphics[width=0.7\linewidth]{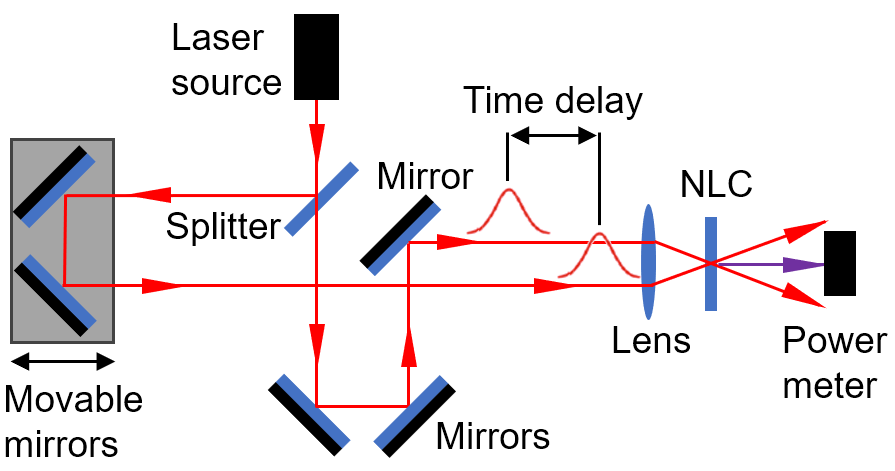}
\caption{Schematic of a noncollinear autocorrelation setup, with beam segments from linear (red arrows) and second-order nonlinear optical responses (purple arrow).\label{fig:nlc_scheme}}
}
\end{figure}

Figure~\ref{fig:nlc_scheme} depicts a typical experimental setup for measuring nonlinear optical responses \cite{trager2012}. 
A laser beam is divided into two paths by a beam splitter. 
After being directed through several mirrors, the two beams are converged by a lens that aligns them to spatially overlap within an NLC.
If the pulses also overlap temporally, their interaction with the NLC leads to the generation of a single sum-frequency beam via a second-order nonlinear response (purple arrow in the figure). 
In addition, each pulse on its own also creates a collinear second-harmonic beam.
The relative time delay between the two split beams can be varied with nanometer precision using a motorized delay stage that controls the position of the attached mirrors (grey rectangle in the figure).
For each time delay, the intensity of the generated beam is captured by a power meter, resulting in a second-order intensity autocorrelation curve that can be studied on a virtual laptop to determine pulse durations.
The shape of the curve is influenced by the pulse shape and pulse duration.
The complexity of such an experiment arises from the need to simulate the spatio-temporal nonlinear optical interactions of multiple interdependent beam segments, while simultaneously considering adjustments by the experimenter in real time.

As shown in Figure~\ref{fig:nlc_scheme}, a beam segment may have more than one direct successor path due to the partial transmittance and reflectance from a beam splitter, or due to nonlinear optical responses, which can lead to complex, nested path branching.
While in linear optics, tracing a path in reverse to its source always reveals exactly one direct sequence of beam segments, a path originating from a nonlinear optical element may have more than one predecessor path leading backwards to a source.
A suitable model must therefore incorporate a self-consistent approach to account for the complex, interdependent path segments created by nonlinear optics and beam splitters, which goes beyond the simple one-to-one mapping of input-to-output segments typically found in linear optics experiments.

Predicting laser paths through heuristics poses significant challenges due to the dependency of each beam segment on its preceding paths, tracing all the way back to a laser source. 
Instead, resolving this dependency necessitates calculating all optical transformations involved in constructing the beam's path.
Furthermore, even minor alterations in any preceding optical element, including the source itself, can significantly alter the beam paths, such as when the laser source is obstructed. 
This sensitivity to physical constraints and interactions underscores the need for detailed optical calculations over heuristic methods.

Although femtoPro aims to create a freely-configurable experimental platform,
the scope of any experiment is constrained by the space needed for optical elements and their manual adjustments. 
Each optical element is mounted on a pedestal.
Considering the dimensions of each pedestal and the virtual hands, with $7~\text{cm} \times 7~\text{cm}$ and $10~\text{cm} \times 20~\text{cm}$ respectively, and assuming adjacent optical elements can share their common interaction space, each element requires a minimum area of $17~\text{cm} \times 17~\text{cm}$, or approximately $0.03~\text{m}^2$, to facilitate comfortable user interactions.
Given these spatial requirements, the optical table can accommodate a few hundred optical elements. However, in most cases, particularly for educational purposes, it is not necessary to reconstruct a full setup with hundreds of optics. Instead, focusing on a specific aspect of an experiment typically suffices, limiting the number of required optical elements to a few tens.
This restriction helps to constrain the extent of detailed optical calculations.
However, even a few optical elements, especially when arranged with NLCs in sequence, can exponentially increase the number of generated beams, posing significant challenges for real-time calculations.

In summary, the simulation must be capable of calculating and rendering beams with their detailed spatio-temporal characteristics in real time as they traverse through various types of freely-configurable optical elements. 
The simulation is required to efficiently manage potentially complex, nested, and branched path structures with interdependent segments that result from optical elements applying linear and nonlinear optical responses to the beams. 
Efficient management of these structures is crucial for dynamic adaptation to user interactions, maintaining the integrity and responsiveness of the experimental environment. 
The following section will detail how we addressed these challenges and met the requirements of the simulation.

\section{Implementation}
\label{sec:implementation}
We chose to model the laser path simulation as a directed graph, naturally representing the one-way flow of laser beams from a laser source through various optical elements to a specific target. 
The full path of a laser beam is divided into segments, with each segment represented as a directed edge, originating from a source node and propagating towards a target node.
Laser sources are modeled as root nodes. They are the initial elements of the simulation, generating beams that travel through the system towards various targets.
Optical elements that transmit and/or reflect beams are depicted as internal nodes.
By means of linear and nonlinear optics calculations they can create outgoing edges based on the properties of incident beams.
All other objects such as measurement devices, beam blockers, the table, or body parts of the virtual avatar, are considered as sink nodes that terminate any further beam propagation.

\begin{figure}[htb]
{
\centering
\includegraphics[width=0.6\linewidth]{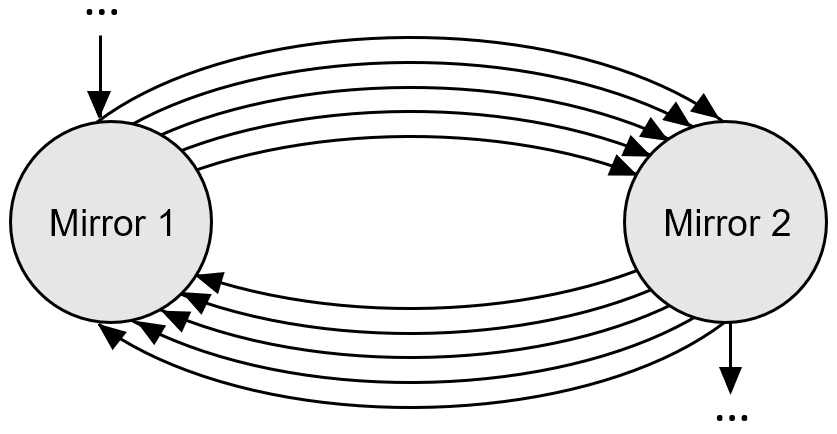}
\caption{Graph representation of Fig.~\ref{fig:cavity_scheme} with black arrows for directed edges and grey circles for nodes.\label{fig:cavity_graph}}
}
\end{figure}

The graph model has characteristics that are unique to a laser path simulation.
Figure~\ref{fig:cavity_graph} shows a graph representation of the finite bouncing scenario discussed in the previous section (see Fig.~\ref{fig:cavity_scheme}) with black arrows for directed edges and grey circles for nodes. 
Despite having the same source and target nodes, directed edges from different round trips cannot be represented as a single edge.
Instead, femtoPro's graph model requires that each round trip be individually modeled with a new set of edges to simulate the entire beam's journey spanning multiple reflections.

\begin{figure}[htb]
{
\centering
\includegraphics[width=0.4\linewidth]{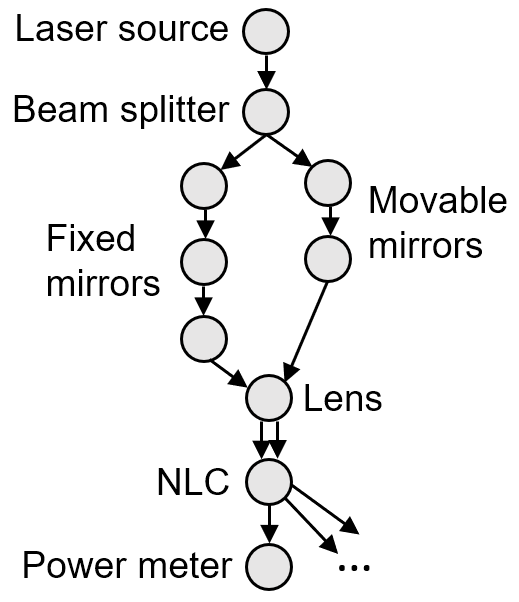}
\caption{Graph representation of Fig.~\ref{fig:nlc_scheme} with black arrows for directed edges and grey circles for nodes.\label{fig:nlc_graph}}
}
\end{figure}

Figure~\ref{fig:nlc_graph} depicts the graph representation of Figure~\ref{fig:nlc_scheme}.
Nodes and edges can be considered ``labeled,'' meaning that they carry detailed optical and geometrical attributes.
These attributes, such as beam direction, index of refraction, or reflectivity of optical media, dictate the flow and behavior of laser beams within the system and thus the graph's topology.

We implemented a dynamic graph model, which supports the modification of nodes as well as the addition and deletion of edges in real time.
The graph's entire topology is revealed only after all laser beams have traversed all relevant optical elements involved in forming their complete pathways through linear and nonlinear optical transformations.
Algorithms for efficient graph traversal, such as breadth-first, depth-first, and A* search, are well-established in both literature and practice \cite{GraphAlgosBook}. 
These methods, however, are generally tailored to static graphs where the entire topology is known in advance, making them unsuitable for a graph that is dynamically generated in real time.
Furthermore, the conventional notion of ``graph traversal'' doesn't precisely align with the concept of beams traversing through optical elements.
In traditional graphs, edge generation and traversal are independent processes, with traversal typically taking place after the graph is completely constructed.
However, in the context of a laser simulation, the generation and traversal of an edge are closely linked, consecutive steps within the ongoing graph generation process.

We therefore developed a custom-tailored path solver algorithm specifically designed to manage the real-time generation and traversal of the graph in our laser simulation (see Algorithm~\ref{alg:PathSolver}).
\FloatBarrier
\captionsetup{type=algorithm}
\captionof{algorithm}{Pseudocode of the laser path solver algorithm, executed each frame. Operations are highlighted in bold and method calls have italic font style.\label{alg:PathSolver}}
\begin{center}
\begin{algorithmic}[1]
\State \textit{ForEachFrame}()
\State \ \ \ \ \textit{DetectChanges}()
\State \ \ \ \ \textit{ProcessNodes}()
\State 
\State \textit{DetectChanges}()
\State \ \ \ \ \textbf{foreach} edge \textbf{in} graph
\State \ \ \ \ \ \ \ \ \textbf{if} \textit{CalcNewLength}(edge) \textbf{!=} edge.length
\State \ \ \ \ \ \ \ \ \ \ \ \ \textit{AddToQueues}(edge.source)
\State \ \ \ \ \ \ \ \ \ \ \ \ \textit{AddToQueues}(edge.target)
\State \ \ \ \ \ \ \ \ \ \ \ \ \textbf{if} edge.type \textbf{==} linear
\State \ \ \ \ \ \ \ \ \ \ \ \ \ \ \ \ \textit{DeletePathStartingAt}(edge)
\State \ \ \ \ \ \ \ \ \ \ \ \ \textbf{else if} edge.type \textbf{==} nonlinear
\State \ \ \ \ \ \ \ \ \ \ \ \ \ \ \ \ \textbf{foreach} sourceOutEdge \textbf{in} edge.source.outEdges
\State \ \ \ \ \ \ \ \ \ \ \ \ \ \ \ \ \ \ \ \ \textit{DeletePathStartingAt}(sourceOutEdge)
\State \ \ \ \ \textbf{foreach} node \textbf{in} graph \textbf{where} isModified \textbf{==} true
\State \ \ \ \ \ \ \ \ \textbf{foreach} edge \textbf{in} node.inEdges
\State \ \ \ \ \ \ \ \ \ \ \ \ \textit{AddToQueues}(edge.source)
\State \ \ \ \ \ \ \ \ \ \ \ \ \textit{DeletePathStartingAt}(edge)
\State \ \ \ \ \ \ \ \ \textit{AddToQueues}(node)
\State 
\State \textit{ProcessNodes}()
\State \ \ \ \ \textbf{while} \textbf{!}\textit{BothQueuesAreEmpty}() \textbf{\&\&} count\textbf{++} \textbf{<} maxCount
\State \ \ \ \ \ \ \ \ \textbf{if} linearQueue.\textit{IsEmpty}()
\State \ \ \ \ \ \ \ \ \ \ \ \ node \textbf{=} nonlinearQueue.\textit{Dequeue}()
\State \ \ \ \ \ \ \ \ \ \ \ \ node.outEdges \textbf{=} \textit{NonlinearTransform}(node.inEdges)	
\State \ \ \ \ \ \ \ \ \textbf{else}
\State \ \ \ \ \ \ \ \ \ \ \ \ node \textbf{=} linearQueue.\textit{Dequeue}()
\State \ \ \ \ \ \ \ \ \ \ \ \ \textbf{foreach} edge \textbf{in} node.inEdges

 \ \ \ \ \ \ \ \ \ \ \ \ \ \ \ \ \ \ \ \ \ \ \textbf{where} edge.needsLinearTransform \textbf{==} true
\State \ \ \ \ \ \ \ \ \ \ \ \ \ \ \ \ node.outEdges \textbf{+=} \textit{LinearTransform}(edge)
\State \ \ \ \ \ \ \ \ \ \ \ \ \ \ \ \ edge.needsLinearTransform \textbf{=} false
\State \ \ \ \ \ \ \ \ \textbf{foreach} edge \textbf{in} node.outEdges
\State \ \ \ \ \ \ \ \ \ \ \ \ edge.needsLinearTransform \textbf{=} true
\State \ \ \ \ \ \ \ \ \ \ \ \ \textit{AddToQueues}(edge.target)
\State 	
\State \textit{AddToQueues}(node)
\State \ \ \ \ linearQueue.\textit{Enqueue}(node)
\State \ \ \ \ \textbf{if} node.type \textbf{==} nlc
\State \ \ \ \ \ \ \ \ nonlinearQueue.\textit{Enqueue}(node)
\State 
\State \textit{DeletePathStartingAt}(edge)
\State \ \ \ \ \textbf{foreach} parentEdge \textbf{in} edge.parentEdges
\State \ \ \ \ \ \ \ \ parentEdge.needsLinearTransform \textbf{=} true
\State \ \ \ \ \textit{DeleteEdgeAndAllItsDescendants}(edge)
\end{algorithmic}
\end{center}
The algorithm relies upon a node queuing approach to iteratively process nodes within the graph, thereby incrementally building and updating its topology.
Processing a node in this context means the generation of new outgoing beams through linear and nonlinear transformation of given incident beams.

In a regular queue data structure, enqueuing adds an element to the end of the queue.
Using a regular queue for the processing of nodes results in the graph being built breadth-first.
This means that the graph is constructed across its full breadth before progressing to the next level of depth.
The use of another queuing priority results in another order of how the graph is constructed.
As an example, always enqueuing nodes at the beginning of a queue would result in a depth-first creation of the graph, where an individual path is fully generated to its maximum depth before other paths are constructed.
In linear optics, since all paths are calculated independently, the order of processing nodes does not have an impact on performance.
However, when considering nonlinear optical elements, calculations could be performed redundantly and unnecessarily, depending on the order that nodes are processed.

\begin{figure}[htb]
{
\centering
\includegraphics[width=0.6\linewidth]{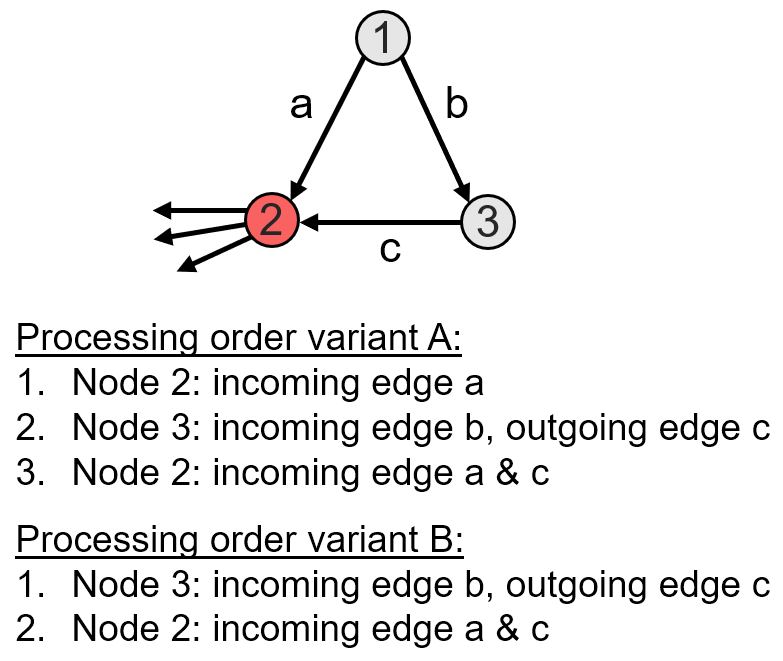}
\caption{Effect of processing order on the number of performed calculations. Node one and three (grey) are linear optical elements and node two (red) is a nonlinear optical element.\label{fig:processing_order}}
}
\end{figure}

Figure~\ref{fig:processing_order} depicts a simple example on how the order of processing nodes impacts the number of (unnecessary) calculations.
Node two is a nonlinear optical element whose incident beams collectively impact the characteristics of any generated outgoing beam.
This means that node two requires all available incident beams to perform a complete nonlinear transformation.
In processing order variant A, however, node two is processed first. 
Since edge a and c are both required input for a nonlinear transformation, but the latter is not yet generated due to node three not yet being processed,
the nonlinear transformation is based upon incomplete input and thus has to be performed twice.
In contrast to this, variant B processed node three first, delivering all required input for the subsequent nonlinear transformation, therefore avoiding
unnecessary and redundant calculations.
To solve this problem, we implemented priority queuing \cite{GraphAlgosBook} to ensure that all linear transformations are processed before any nonlinear transformations.
This guarantees that a nonlinear optical element waits for complete input of linearly transformed beams before it performs a nonlinear transformation.
However, if an experimental setup contains more than one nonlinear optical element and those elements depend on each other's input of nonlinearly transformed beams, there is still a chance that a nonlinear transformation may be performed redundantly with incomplete input.
This issue is not yet solved in femoPro, since experimental setups rarely require more than one nonlinear optical element.
A potential solution could involve analyzing the pathways previously generated by linear transformations to determine an efficient processing order of multiple dependent nonlinear transformations.

After having outlined the queuing approach for processing nodes, we can now dive deeper into the pseudocode listing of the laser path solver algorithm.
Each frame, the method DetectChanges (Lines~5--19) and then ProcessNodes (Lines~21--33) is executed.
As first step of the change-detection method, we test whether the length of any edge inside the graph has changed (Line~7).
This corresponds to a beam being obstructed, e.g., by the hand of an avatar.
If the length of an edge has changed, its source and target node are enqueued for further processing (Lines~8--9).
We utilize two separate first-in-first-out queue data structures: one for nodes that require linear transformations (Line~36) and another for nodes additionally requiring nonlinear transformations (Line~38), with each queue ensuring that all its entries are unique, meaning no node is enqueued more than once. 
This approach allows us to prioritize processing all linear transformations before processing any nonlinear transformation, which we will explain in more detail later.
Since the edge and all of its descendants are outdated due to the introduced change, they have to be deleted from the graph (Line~11).
A key requirement of the path-deletion method (Lines~40--43) is that each edge must maintain references (i.e., pointers) to its adjacent path segments---that is, the preceding parent edges and the following child edges.
Any given path segments can therefore be consecutively deleted from the graph by following the pointers that connect these edges.
For brevity and simplicity, the management of those pointers is not depicted in the pseudocode.
Before the deletion of respective path segments (Line~43), their unchanged parent edges must be labeled as requiring a linear transformation (Lines~41--42). This is necessary, because otherwise, the deleted path segments would not be rebuild later, since linear 
transformations are only performed for labeled edges (Lines~28--29) to avoid redundant calculations of unchanged edges and their subsequent paths.
In summary, adding the edge's source and target nodes to respective processing queues (Lines~8--9) as well as labeling all parent edges (Line~42) guarantees that the changed edge and all its potential descendants are incrementally rebuilt later (Lines~21--33).

In the special case where the edge was generated by a nonlinear transformation, rather than just deleting the edge's path, all paths originating from the NLC have to be deleted (Lines~12--14).
This is because all beams incident on an NLC can collectively impact the characteristics of all nonlinear outgoing beams, rendering the state of all outgoing beams outdated. 
Furthermore, since an NLC also adheres to the principles of linear optics, all beams incident on the NLC have to be labeled to require a linear transformation (Line~42).
Because a nonlinear transformation requires all incident edges as combined input (Line~25), adopting a labeling approach similar to the one used for selectively performing linear transformations is not feasible.

After having dealt with interrupted beam segments, all nodes of the graph have to be checked for any modifications (Lines~15--19).
This could, for example, be a change in geometrical attributes (such as a change in the angle of a mirror) or optical characteristics (such as the change of pulse duration at the laser source).
If a node is modified, not only the node itself (Line~19) but also all source nodes of all of its incoming edges have to be enqueued (Line~17). 
This is necessary because a modification to a node, such as a change in position, also impacts properties of the node's incident beams, like their path length, which means these beam segments must be recalculated as well.
Since all edges incident on a modified node are thus outdated, their parent edges have to be labeled to require a linear transformation and their subsequent paths have to be deleted (Line~18).
After all source nodes of all incoming edges are enqueued, the modified node itself is also added to the queues.
This sequence guarantees that parent edges are processed before any child edge, avoiding unnecessary dequeuing iterations.

After the queues are filled with all nodes affected by change and respective edges are labeled for requiring a linear transformation, the method for processing the queues is executed (Lines~21--33).
As long as any of the queues have entries and the maximum iteration count is not reached, the algorithm continues to dequeue and process a node within a while loop (Line~22).
If the linear transformation queue is empty, a nonlinear transformation is performed on a node dequeued from the nonlinear transformation queue (Lines~23--25).
Since in this case, all incident beams may collectively influence the characteristics of any outgoing beam, they all have to be passed into the transformation method.
This includes both edges that require and do not require any transformation (Line~25).
The underlying pulse transformation logic is interchangeable and can be tailored to the specific needs of the application.
Thus, the algorithm can be thought of as a flexible infrastructure for delivering various physics-based calculations with required accuracy and performance.
After the transformation process, the newly generated outgoing edges are stored inside the node's respective attribute (Line~25).

As long as the linear transformation queue has entries, new outgoing edges are generated through linear transformations (Lines~26--29).
This is only performed for incoming edges that are labeled to require a linear transformation, effectively avoiding redundant calculations of incident edges whose
properties have not changed compared to the last frame (Line~28).
After its linear transformation, the respective edge has to be labeled to no longer require a linear transformation (Line~30).
This is important to prevent unnecessary recalculations in future loop iterations, i.e., when the same node is enqueued again due to other subsequent paths traversing it. 
As a final step of the processing loop, each newly generated outgoing edge is labeled and the corresponding target node is enqueued for further processing (Lines~31--33).
The cascading, repetitive process of dequeuing, transforming and enqueuing guarantees that each node is visited and its connections are fully explored, thereby dynamically building the graph according to given optical laws.
The overall optics simulation converges when both the linear and nonlinear transformation queues are empty or the maximum iteration count is reached. 
It restarts every frame to account for any changes introduced by user interaction, ensuring that the state of the simulation is correct and up-to-date.

Since the path solver algorithm acts as a framework for various physics-based calculations whose performance cost may greatly vary based on desired accuracy,
it is fundamental for scalability that the path solver itself performs those calculations in a redundant-free manner.
In summary, by performing linear transformations only selectively for edges that are affected by change and by using priority queuing to handle nonlinear transformations, we eliminated any redundant and unnecessary calculations.
This prevents wasting computational resources on aspects of the system that remain unchanged or have incomplete input, thereby making the overall simulation more efficient.
Moving forward, we will refer to this approach as the selective updating strategy.

\begin{figure}[htb]
{
\centering
\includegraphics[width=0.6\linewidth]{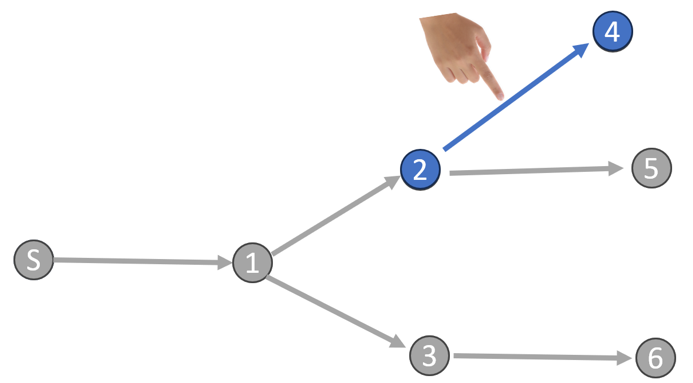}
\caption{Effect of beam interruption by the hand of an avatar. Only the blue edge is recalculated.\label{fig:selective_path}}
}
\end{figure}

Figure~\ref{fig:selective_path} depicts an example on how the selective updating strategy positively affects the efficiency of the simulation. 
The depicted blue beam is interrupted between node two and four by the hand of an avatar.
Since in the example, the interrupted edge does not have any descendant edges, only this edge is recalculated.
Every other edge remains unchanged and therefore no redundant calculations are performed.

Efficient memory management techniques, although not included in the pseudocode, are crucial for the performance of the path solver's implementation.
A naive approach would be to allocate new memory for each beam segment that is newly generated by linear and nonlinear transformations within the node processing loop.
Without deallocating the outdated beam segments from the previous frame, which have become irrelevant due to introduced changes, this would inevitably lead to memory overflow.
To mitigate this issue, software frameworks such as game engines, including Unity, employ a garbage collection (GC) system that automatically frees up resources no longer in use by the application \cite{gcInUnity2024}.
However, this process can temporarily halt other operations, leading to noticeable drops in the frame rate.
These drops are particularly detrimental in a VR context because they can disrupt the immersive user experience, potentially causing severe discomfort such as motion sickness \cite{effectOfFramerate2023}.

To ensure consistent and smooth performance, we needed a method to manage memory more efficiently than relying solely on GC, which led to the implementation of an extensive pooling system \cite{Gregory2009}. 
A pooling system preallocates a block of memory for objects that are frequently created and destroyed, in this case, the beam segments of the simulation. 
Instead of allocating new memory for each newly generated segment, segments are taken from this preallocated pool, and outdated segments from the previous frame are returned to the pool, making them available for reuse.
When the pool reaches its capacity, it automatically expands by a predetermined size to accommodate more beam segments.
It is important to note that pooling is helpful not only for managing beam segments but also for handling temporary data structures within physics calculations, especially those with a potentially large sample size.
Initially, we utilized the Math.NET framework for data structures and operations \cite{mathNetWebsite}, but it proved unsuitable as it allocated new memory for each computational operation rather than modifying variables in place.
Since none of the solutions on the market met our specific needs for efficient real-time data processing, we developed our own pooled data structures that support in-place operations.

%
\section{Performance Analysis}
\label{sec:analysisMain}
%
%
In Section~\ref{sec:analysisInGeneral}, we determine the general computational cost of the simulation and assess its scalability,
while Section~\ref{sec:analysisOfExperiments} focuses on evaluating the resource demands of specific experimental setups implemented in the current version of femtoPro. Section~\ref{sec:analysis_practicalConsiderations} explores the practical implications of algorithmic performance on user experience.

%
\subsection{General performance analysis}
\label{sec:analysisInGeneral}
%
%
We first perform a memory analysis of the path solver algorithm.
An experimental setup can have a variable number of optical elements and beam segments.
Each of those elements and segments has a fixed number of attributes to define their characteristics.
Depending on its data type, storage of an attribute requires a certain amount of memory (in Bytes).
As an example, a double-precision floating-point number requires 8~Bytes \cite{sizeOfDouble}. 
To determine the total memory cost of an optical element or beam segment, we sum up the type-specific cost of each of its attributes.
Besides simple types such as integers and boolean flags, beam segments also have attributes of complex type, such as arrays.
For instance, a beam's frequency and wavelength values are stored in arrays with a specific sampling size.
To determine the memory cost of those complex types, we simply have to multiply the per-element cost (e.g., 8~Bytes for double-precision floating-point numbers)
with the sampling size.
Conclusively, we calculate the memory requirements, $M$, of an experimental setup as follows
\begin{center}
\begin{subequations}
        \begin{equation}
            M = e  C_\text{e} + b  C_\text{b},\label{eq:totalMemory}
        \end{equation}
        \begin{equation}
            C_\text{e} = \sum_{s=1}^{N} \left(A_{s}  C_\text{s}\right),\label{eq:opticalElementMemory}
        \end{equation}
        \begin{equation}
           C_\text{b} = \sum_{s=1}^{N} \left(A_{s}  C_\text{s}\right) + \sum_{c=1}^{N} \left(A_{c} C_\text{c}  S\right),\label{eq:beamMemory}
        \end{equation}
\end{subequations}
\end{center}
including 
the number of optical elements $e$,
the summed cost for each optical element $C_{\text{e}}$,
the number of beams $b$,  
the summed cost for each beam $C_b$,
the attribute count $A$ for each simple ($s$) and complex ($c$) type, 
the cost $C$ associated with each type,
and the sampling size $S$ for complex types.
Due to the implementation of an extensive pooling system and in-place operations for large data structures (see Section~\ref{sec:implementation}), the memory footprint
of the simulation is limited to the outcome of Equation~\ref{eq:totalMemory}.

Conservatively estimated, the total memory budget for applications on a mobile VR device is at least 1~GB \cite{metaQuestRamArticle}. 
From this available memory, each optical element requires approximately 4~KB, as calculated by inputting respective values into Equation~\ref{eq:opticalElementMemory}.
Over 260,000 optical elements would be required to max out femtoPro's available memory, a number that is far above the limit of several hundred optical elements imposed by physical and spatial constraints (see Section~\ref{sec:requirements}).
For the currently implemented experimental objectives, a sampling size of $S=512$ is sufficient.
As a result, by applying Equation~\ref{eq:beamMemory}, we calculate that each beam segment requires approximately 16~KB of memory.
Thereof, 98~\% is allocated for array structures proportional to the sampling size, representing the largest chunk of pulse data.

With a sampling size of $S=512$, about 65,000 beams would be required to reach the memory limit.
This number of beams could theoretically be reached by configuring mirrors to form a cavity (compare Fig.~\ref{fig:cavity_scheme}). 
In practice, however, the number of beams generated typically remains well below 1,000 in all other scenarios.
Additionally, the simulation incorporates a safeguard that terminates the path solver algorithm after a specified number of laser light bounces between optical elements (see Algorithm~\ref{alg:PathSolver}, Line~22), preventing memory overflow.
In conclusion, thanks to an efficient algorithm and the decently large available RAM capacity, memory costs are not a concern in practice.

%

Next, we examine the computational performance of the path solver algorithm and the simulation calculations.
Written as single-threaded C\# code, we narrow the scope of analysis to the CPU's workload for executing the algorithm.
We exclude external factors and impacts from the broader system context on the simulation, such as bottlenecks in graphics or memory processing that could indirectly affect CPU performance.
Although beam rendering could be considered as part of the simulation, it is not directly related to the simulation's code and therefore is disregarded.

The sampling size $S$ for electric field representations significantly affects CPU performance, since certain operations must be performed for each element in respective field arrays.
To simplify the analysis of computational performance, we assume that the sampling size is constant.

Given the extensive ways in which optical elements can be manipulated, generalizing the simulation's overall performance cost is challenging without considering a specific experimental alignment with fixed characteristics.
For example, when a user redirects a laser beam with a mirror, the execution time within the current frame can vary significantly---ranging from very low, such as when the redirected beam is terminated by a beam blocker, to very high, such as when the beam forms a cavity. 
Consequently, the expected cost of the simulation at any given time depends on the outcomes of interactions determined by the previous iteration of the simulation. 
To generally understand performance costs despite these dependencies, we limit our examination to modifications that do not alter the graph's topology, such as changes to the pulse duration within a laser source.

Under the specified constraints, the overall execution time $C_{\text{total}}$ of the simulation within a frame is given by
\begin{equation} \label{eq:cost_total}
    C_{\text{total}} = C_{\text{sel}} + C_{\text{lin}} + C_{\text{nonlin}},
\end{equation}
where $C_{\text{sel}}$ is the time spent for maintaining the selective updating strategy,
and $C_{\text{lin}}$ and $C_{\text{nonlin}}$ represent the time spent for linear and nonlinear transformations, respectively.

The selective updating strategy is responsible for monitoring whether any of the nodes or edges of the graph have been modified between consecutive frames (see Alg.~\ref{alg:PathSolver}, Lines~5--19).
The execution time for $C_{\text{sel}}$ can therefore be calculated as
\begin{equation} \label{eq:cost_sel}
    C_{\text{sel}} = b  C_{\text{b}} + e  C_{\text{e}},
\end{equation}
where $b$ and $e$ represent the total number of beams and optical elements, respectively, that are currently present in the simulation.
$C_{\text{b}}$ is the execution time per beam, 
and $C_{\text{e}}$ the execution time per optical element.

The execution time for all linear transformations in a frame, $C_{\text{lin}}$, is determined by
\begin{equation} \label{eq:cost_lin}
    C_{\text{lin}} = \sum_{i = 1}^{N} \left[b_{i}  (C_{\text{reflect}} + C_{\text{transmit}})\right],
\end{equation} 
where $b_i$ represents the number of modified incident beams on each optical element $i$ out of $N$ total optical elements that require a linear transformation.
$C_{\text{reflect}}$ and $C_{\text{transmit}}$ denote the time spent per beam for calculating reflection and transmission, respectively.
It is noteworthy that linear interactions also occur within modified nonlinear optical elements, so the associated costs must also be applied to these elements.

Currently, femtoPro supports only second-order nonlinear interactions, with third-order response still under development.
Second-order phenomena, such as sum-frequency generation (SFG) and second-harmonic generation (SHG), inherently involve interactions between beams.
These can occur between two distinct beams, as in SFG, or within a single beam, as in SHG, where the frequency of the resulting beam is twice that of the original.
The execution time for these second-order responses can be expressed by
\begin{equation} \label{eq:cost_nonlin}
    C_{\text{nonlin}} = \sum_{i = 1}^{N} \left(\frac{b_i  (b_i + 1)}{2}  C_{\text{pair}}\right),
\end{equation}
where $b_i$ represents the number of incoming beams for the $i$-th element out of $N$ total nonlinear optical elements that are modified
and thus require a nonlinear transformation.
The term $\frac{b_i  (b_i + 1)}{2}$ applies the combinatorial concept of ``multichoose'' to calculate the total number of unique pairwise combinations of incident beams $b_i$, including self-interactions for SHG. 
$C_{\text{pair}}$ is the cost associated with processing each of these pairwise combinations.

Although the derived equations are based on restrictive theoretical assumptions, thus offering a limited view of potential performance demands in practice, they remain useful for approximating the computational demands of any planned experimental setup.
This approach reduces the need for extensive benchmarking, as will be detailed in Section~\ref{sec:analysisOfExperiments}.

Due to the selective updating strategy, the execution time may vary significantly depending on what node or edge of the graph's topology is modified.
We therefore define a best-, average-, and worst-case scenario, each reflecting the different impacts specific modifications have on processing time.
The best-case scenario occurs when none of the graph's edges or nodes have changed.
Under these conditions, computational costs are limited to maintaining the selective updating strategy, resulting in both $C_{\text{lin}}$ and $C_{\text{nonlin}}$ being zero.
The worst-case scenario arises when the laser source is modified, necessitating the recalculation of all edges within the graph.
Under the previously discussed assumption that modifications to nodes do not alter the graph's topology, the execution times for both $C_{\text{lin}}$ and $C_{\text{nonlin}}$ reach their maximum. 

We define the average-case execution time, $C_{\text{avg}}$, as
\begin{equation} \label{eq:cost_avg}
    C_{\text{avg}} = \sum_{i}^N (p_i  C_{\text{total},i}),
\end{equation}
where $p_i$ represents the percentage of frames in which the user modifies a specific node $i$ out of $N$ total nodes
and $C_{\text{total},i}$ denotes the overall cost associated with modifying a specific node. 
Each $p_i$ is calculated as a fraction of the total number of frames containing a node interaction.
Consequently, the sum of all $p_i$ equals 1.
We again assume that a modification to a node does not alter the graph's topology.
For simplicity, we only consider node modifications (Alg.~\ref{alg:PathSolver}, Lines~15--19) and not individual edge interruptions (Alg.~\ref{alg:PathSolver}, Lines~6--14).
This approach is adequate for a conservative estimate because modifying a node has the same impact on the simulation as interrupting all of its incident beams.
In preparation for the analysis in Section~\ref{sec:analysisOfExperiments}, we will define the modification of a specific node $i$ out of $N$ total nodes as one specific node interaction event among all possible node interaction events of an experimental setup.
Therefore, $C_{\text{avg}}$ represents the weighted average execution time of all these events, with weights corresponding to the percentage of frames in which each event occurs.

Runtime complexity is a fundamental concept in computer science that describes how the computational time requirements of an algorithm change as the size of the input data grows. 
It provides a way to quantify the efficiency of algorithms and predict their performance under different conditions. 
The notation used to express runtime complexity includes ``Big Omega'' ($\Omega$), which represents the lower bound or best-case scenario, and ``Big O'' ($O$), which denotes the upper bound or worst-case scenario \cite{introToComplexity2022}.

In the context of determining the simulation's runtime complexity, the cost values associated with each optical element (i.e., $C_{\text{e}}$), beam (i.e., $C_{\text{b}}$, $C_{\text{reflect}}$, and  $C_{\text{transmit}}$) and beam comparison (i.e., $C_{\text{pair}}$) are considered constant.
This is because the operations per element do not vary with changes in the input data size (i.e., the number of optical elements, beams, and beam comparisons). 
In the specified best-case scenario, runtime complexity is solely dependent on the term of $C_{\text{sel}}$ due to $C_{\text{lin}}$ and $C_{\text{nonlin}}$ both being zero, resulting in $\Omega(b+e)$.
This linear relation indicates efficient scaling with respect to the number of beams $b$ and optical elements $e$.

The overall runtime complexity of an algorithm is determined by the fastest-growing term. 
In the worst-case scenario for our simulation, the quadratic term from the nonlinear transformation, $C_{\text{nonlin}}$, dominates, resulting in a complexity of $O(b^2)$.
However, this quadratic worst-case complexity oversimplifies the nuanced impact of the distribution of incident beams $b$ across multiple nonlinear optical elements.
While, for example, the complexity for one NLC with five incoming beams is quadratic ($5^2 = 25$), it is only linear for five NLCs with each having one incoming beam $\left((5)  (1^2) = 5\right)$.
Hence, the runtime complexity can be significantly less than $O(b^2)$, depending on the beam distribution among nonlinear elements.
Nonetheless, the presence of quadratic complexity in the worst-case scenario highlights the challenges inherent to nonlinear transformations.

Ideally, typical experimental setups that reflect common practical configurations should be extensively benchmarked on each specific target platform to ensure they meet the real-time requirements for a smooth user experience.
Continuously building the application, installing it on VR devices, running the simulation over thousands of iterations, and exporting as well as interpreting gathered benchmark data can be cumbersome and time-consuming.
Instead, Equations~\ref{eq:cost_total}--\ref{eq:cost_nonlin} can be integrated into an algorithm to automatically calculate worst-case computational cost inside a game engine's level editor.
This allows for estimating the execution time for various target platforms, without running the application, streamlining the workflow for creating and adjusting predefined experimental setups.
This only requires that entity-specific costs (i.e., costs per optical element, beam, or beam comparison) have been determined on each desired target platform.

In this context, Table~\ref{tab:perelement} presents the execution times per entity.
All execution time measurements of Section~\ref{sec:analysisInGeneral} and \ref{sec:analysisOfExperiments} were gathered using the release version 0.7.0 of femtoPro for Android on the Pico~3 VR headset \cite{femtoProWebsite,pico3Website}.
We employed the stopwatch feature of C\# \cite{stopwatchClass} to mark specific parts of the simulation code for time measurements. 
For each term listed in the first column, we calculated the average execution time across 1,000 runs to minimize variations caused by other system components competing for CPU time (see column two).
Notably, the execution time for $C_{\text{pair}}$, at approximately 113 \textmu s, significantly exceeds other per-entity costs, which all remain below 60 \textmu s.

\begin{table}[htb]
\centering
\caption{Average execution times per entity (i.e., optical element, beam, or beam comparison).\label{tab:perelement}}
\begin{tabular}{|p{2.8cm}|p{3cm}|}
\hline
Term & \raggedleft Time measured in \textmu s \tabularnewline \hline
$C_{\text{e}}$        & \raggedleft{52.9}  \tabularnewline \hline
$C_{\text{b}}$        & \raggedleft{6.9}  \tabularnewline \hline
$C_{\text{reflect}}$  & \raggedleft{6.8}   \tabularnewline \hline
$C_{\text{transmit}}$ & \raggedleft{21.6}  \tabularnewline \hline
$C_{\text{pair}}$     & \raggedleft{112.8} \tabularnewline \hline
\end{tabular}
\end{table}

In summary, due to their nonlinear complexity and their high per-beam-comparison cost $C_{\text{pair}}$, nonlinear interactions significantly increase the difficulty of handling large numbers of beams compared to experiments involving only linear optics.
For a more detailed analysis, Section~\ref{sec:analysisOfExperiments} will explore the practical impacts of linear and nonlinear transformations on computational costs in typical experimental setups implemented in femtoPro.

%
\subsection{Experiment-specific performance analysis}
\label{sec:analysisOfExperiments}
Although femtoPro's freely configurable and scalable lab environment allows for a wide range of custom experimental configurations, we included various predefined scenarios, or setups, designed for educational purposes that represent typical use cases aligned with standard curricula of established textbooks\cite{trager2012,diels2006,wolfgang2013}.
The layout of every experiment is tailored to specific educational objectives, spanning from basic (e.g., adjusting optical elements) to more advanced tasks (e.g., measuring nonlinear autocorrelation, see Fig.~\ref{fig:nlc_scheme}).

\begin{figure*}[ht]
{
\centering
\includegraphics[width=1\textwidth]{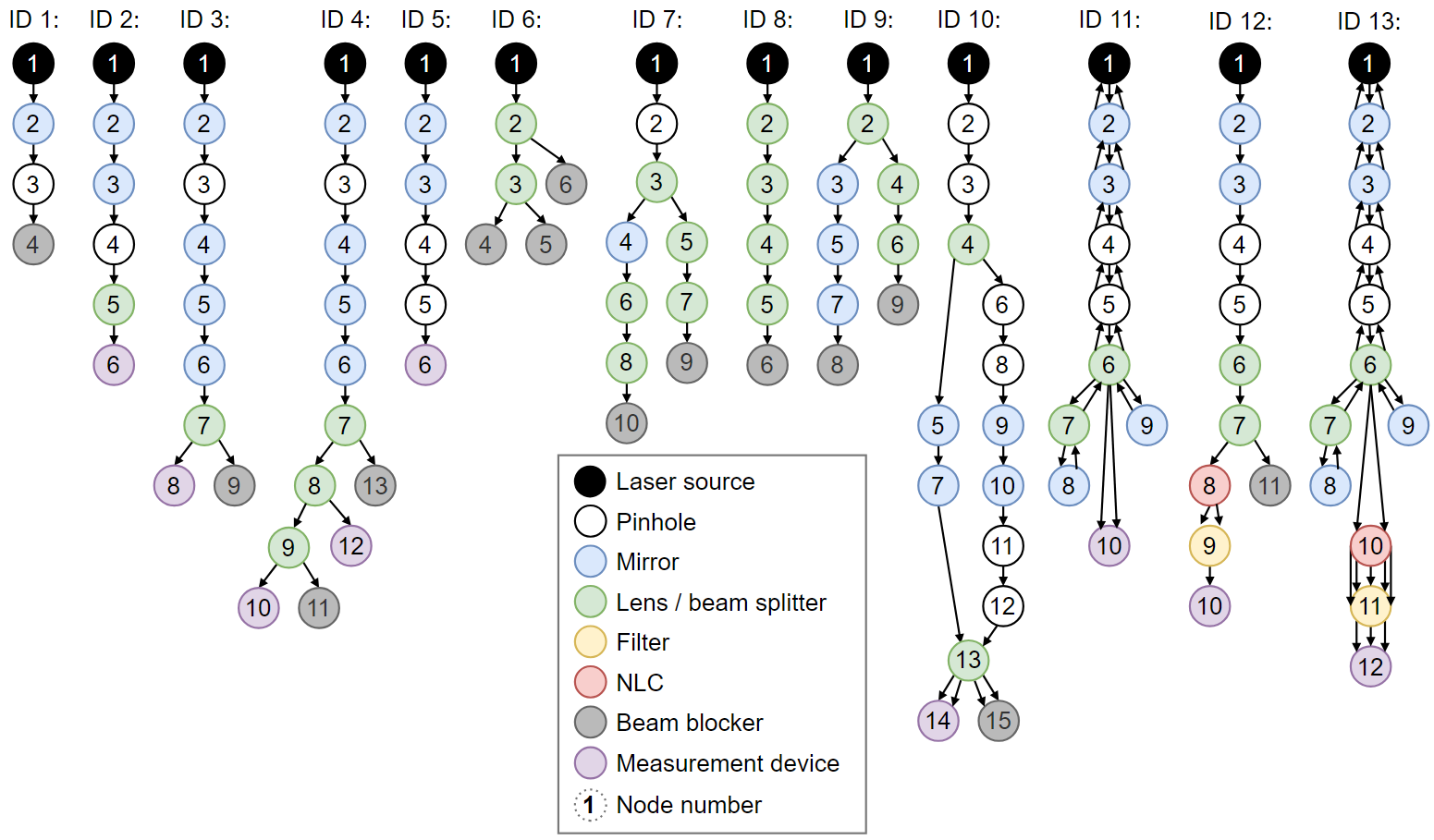}
\caption{Graph representation of experimental setups currently implemented in femtoPro. \label{fig:experimentgraphs}}
}
\end{figure*}
\begin{figure}[htb]
{
\centering
\includegraphics[width=1\linewidth]{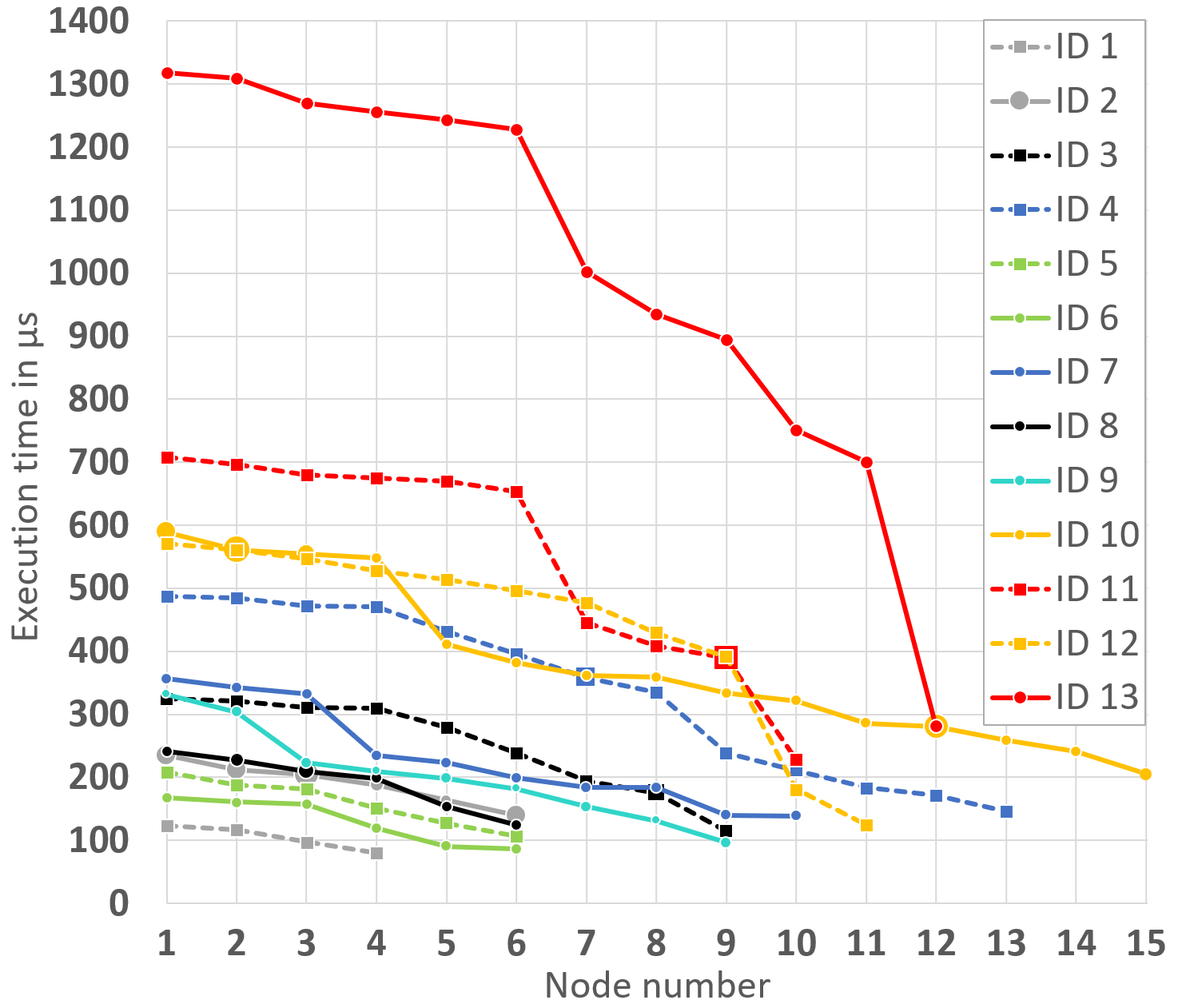}
\caption{Execution times for each interaction event of all implemented experimental setups. Each experiment is represented by a colored line, with dots of the same color marking the execution times for all of its events.\label{fig:allcase_benchmark}}
}
\end{figure}

For a clear understanding of the scope and details of each implemented experiment, Figure~\ref{fig:experimentgraphs} depicts their graph representations.
In these graphs, directed edges are shown as black arrows, and nodes are indicated by numbered circles, with specific colors representing their types.
Each experiment is labeled with a unique ID.
Experiment IDs and node numbers are consistently used across Table~\ref{tab:benchmark} and Figure~\ref{fig:allcase_benchmark}, which will both be introduced later, allowing for easy cross-referencing between diagrams and evaluation data.

Given that the memory usage remains well within available capacity (see Section~\ref{sec:analysisInGeneral}) and it is unlikely that any experimental setup will approach this limit, the following analysis will primarily focus on time efficiency of CPU operations.
For virtual reality experiences, maintaining a minimum frame rate of 120~fps is crucial for minimizing nausea and discomfort \cite{effectOfFramerate2023}.
However, since most currently available mobile VR headsets, including those targeted by femtoPro, only support up to 90~fps, we establish this as our minimum acceptable frame rate, corresponding to a time budget of ca. 11~ms per frame ($\frac{1000~\text{ms}}{90}=11~\text{ms}$).

For mobile platforms, game engines such as Unity advise reserving 35~\% or more of the available frame time for CPU idling. 
This approach aids in controlling device heat and preserving battery life \cite{frameIdleBudget2024}.
At 90~fps, this equates to approximately 4~ms of idle time per frame.
Based on performance profiling on our target platform, we additionally set aside a sufficiently large buffer of 3~ms for system components, such as sensory input processing.
As a result, the maximum time budget per frame that femtoPro can allocate to the simulation code is approximately 4~ms.

We list the worst-case execution time for each of the currently implemented experiments in Table~\ref{tab:benchmark}.
Experiment IDs are listed in column one, with their corresponding names, reflecting their content and objectives, provided in column two.
The third column lists the worst-case execution times measured on the target platform.
These times were obtained by programmatically marking the root node of each experiment as ``hasChanged'' and calculating the average execution time across 1,000 runs.
The fourth column lists the computational time calculated via Equations~\ref{eq:cost_total}--\ref{eq:cost_nonlin}.
In most of the experiments, the discrepancy between the measured and calculated execution time was below 10~\%, while 3 out of 13 experiments had a discrepancy between 10~\% and 20~\% (see Col. 5).
Given the complex interplay of various factors affecting execution time, coupled with the aggregation of minor inaccuracies in measuring average execution times per entity, the discrepancies of the formulaic calculations are in an acceptable range for roughly estimating real-world computational demands of experimental setups.

All experiments are well within the available time budget of 4,000~\textmu s.
However, experiments 11 and 13 exhibit relatively high execution times compared to other setups.
This is primarily because reflections not relevant to the mission objectives, originating from node 6, are fed back into the system (see Fig.~\ref{fig:experimentgraphs}).
This leads to pulses moving towards the root node 1, resulting in a significant increase in the number of edges---23 and 34 for experiments 11 and 13, respectively---that require computational processing.
As demonstrated by this concrete example, the risk of beams reflecting off surfaces and unintentionally re-entering the system is an inherent challenge in optical systems. 
Such reflections can prompt recalculations in unintended areas or even the entire system, compromising the selective update strategy's effectiveness and thus the average-case performance, as will be discussed later.

\begin{table*}[htb]
\centering
\caption{Worst-case performance of implemented experiments.\label{tab:benchmark}}
\begin{tabular}{|r|l|p{2cm}|p{2cm}|p{1.5cm}|}
\hline
ID & Name & Time measured in \textmu s & Time calculated in \textmu s & Delta time in \% \\ \hline
1 & Lock Optical Devices           & \raggedleft 123.4  & \raggedleft 129.5  & \raggedleft 4.9   \tabularnewline \hline
2 & Alignment on Powermeter        & \raggedleft 234.6  & \raggedleft 227.6  & \raggedleft -3.0  \tabularnewline \hline
3 & Lenses                         & \raggedleft 324.8  & \raggedleft 296.2  & \raggedleft -8.8  \tabularnewline \hline
4 & Beam Splitters                 & \raggedleft 487.6  & \raggedleft 454.9  & \raggedleft -6.7  \tabularnewline \hline
5 & Alignment on Irises            & \raggedleft 207.6  & \raggedleft 227.6  & \raggedleft 9.7   \tabularnewline \hline
6 & Kepler Telescope I             & \raggedleft 167.2  & \raggedleft 190.0  & \raggedleft 13.6  \tabularnewline \hline
7 & Kepler Telescope II            & \raggedleft 356.2  & \raggedleft 340.5  & \raggedleft -4.4  \tabularnewline \hline
8 & Galilei Telescope              & \raggedleft 240.6  & \raggedleft 219.5  & \raggedleft -8.8  \tabularnewline \hline
9 & Reflecting Telescope           & \raggedleft 331.5  & \raggedleft 273.4  & \raggedleft -17.5 \tabularnewline \hline
10 & Mach–Zehnder Interferometer   & \raggedleft 589.2  & \raggedleft 643.1  & \raggedleft 9.1   \tabularnewline \hline
11 & Linear Dispersion Measurement & \raggedleft 707.3  & \raggedleft 809.8  & \raggedleft 14.5  \tabularnewline \hline
12 & Second-Harmonic Generation    & \raggedleft 570.4  & \raggedleft 529.9  & \raggedleft -7.1  \tabularnewline \hline
13 & Nonlinear Autocorrelation     & \raggedleft 1318.3 & \raggedleft 1422.3 & \raggedleft 7.9   \tabularnewline \hline
\end{tabular}
\end{table*}

To comprehensively assess the impact of the selective updating strategy on performance across all interaction scenarios---ranging from worst- to best-case---a detailed analysis was carried out. 
The results are presented in Fig.~\ref{fig:allcase_benchmark}.
As discussed in Section~\ref{sec:analysisInGeneral}, we define the modification of a specific, numbered node among the $N$ total nodes of an experiment as an interaction event.
We measured the execution time for all interaction events of each experiment.
In other words, we measured the computational effort caused by manipulating the optical elements, one after another, in an otherwise fixed experimental setup. 
These measurements include both the execution time for optical calculations and maintaining the selective updating strategy.

In this context, a dot in the graphs represents the execution time associated with an interaction event, i.e., the modification of a specific node of a specific experiment.
To visually distinguish between the outcomes of different experimental setups, all interaction events belonging to a specific experiment are connected by a distinct line.
Each line is labeled with the respective experiment ID, consistent with Table~\ref{tab:benchmark} and Figure~\ref{fig:experimentgraphs}.
The vertical axis represents the execution time in \textmu s per frame.
The horizontal axis lists all node numbers for each experiment, corresponding to those shown in Fig.~\ref{fig:experimentgraphs}, with higher node numbers indicating lower execution times.
This means that nodes on the horizontal axis range from the root node (e.g., laser source) as the leftmost data point, representing the cost for the worst-case scenario, to the sink nodes (e.g., beam blockers or measurement devices) as the rightmost data points, approaching best-case scenario costs.
For example, the first data point on the uppermost line indicates the worst-case execution time of 1319 \textmu s for experiment 13 (see Table~\ref{tab:benchmark}), resulting from the modification of node 1 in the corresponding experiment graph, which is, according to Fig.~\ref{fig:experimentgraphs}, the laser source.
As indicated by Figure~\ref{fig:allcase_benchmark}, all currently implemented experiments adhere to the tight time budget of mobile VR devices (here \textasciitilde4~ms) for a smooth real-time experience.
However, for experiments 11 and 13, the costs remain high across most events before rapidly decreasing near their rightmost data points.
The high cost at the first data point is caused by the relatively high number of edges that require optical transformations (see Fig.~\ref{fig:experimentgraphs}).
After the first data point, the costs stay on a relatively high plateau due to reflections in the lower parts of the graph being redirected back towards the laser source.
This redirection leads to changes in the lower parts of the graph requiring also the recalculation of the upper parts, undermining the efficiency of selective updating.
Execution time significantly decreases only for those nodes which are not involved in any redirection of edges back to the source, such as the measurement device in experiment 13.

Data points of experiment 12 indicate that, although the presence of an NLC typically increases execution time, its impact on overall performance is less pronounced than the combined effect of total edge count and edges being redirected to the source.
Analysis of the collected data identifies these edge dynamics as the primary bottlenecks to the path solver's efficiency.
This is confirmed by experiment 11, whose execution times are higher than experiment 12 across most of its interaction events, although there is no NLC involved.
In experiments 7, 9, and 10, there is a relatively sharp decline in execution time after the first few data points.
This is due to the path being split early in its progression. 
Thanks to selective updating, interactions affecting only one branch do not require the recalculation of other branches.
Consequently, the efficiency benefits of nodes subsequent to a path split grow with the number and size of the branches that remain unaffected.

In summary, the selective updating strategy significantly reduces execution times, often well below the worst-case scenario costs, for experiments involving both linear and second-order nonlinear responses. 
However, the efficiency of this strategy is limited by edges that are redirected from lower parts of the graph back to the source. 
Despite this limitation, all experimental setups are well within the execution time budget for a smooth user experience. 
The remaining free budget of at least 50\% provides a substantial buffer to manage any unintended edge dynamics as outlined above, with
a predefined maximum iteration count (see Alg.~\ref{alg:PathSolver}, Line~22) being the upper boundary of light bounces for extreme scenarios such as cavities.

%

\subsection{Considerations on user experience}
\label{sec:analysis_practicalConsiderations}

To enhance safety, femtoPro encourages established practices for safe laser handling.
Safety protocols dictate that the coarse alignment of an optical element is performed without any beams emitting from its surface.
This can be achieved by blocking all beams incident on the optical element being adjusted---using either a stationary beam blocker or a handheld card.
This method allows for the incremental construction of the experiment's layout, minimizing the risk of unintended interactions and exposure to hazardous laser radiation. 
An optical element that is not hit by any beams does not require optics calculations during its alignment. 
Consequently, updating modified parts of the graph occurs in significantly fewer frames, e.g., when a beam blocker is removed from a beam's path
revealing all alignment modifications made while the beam was blocked.
Therefore, this practice complements the selective updating strategy, boosting simulation efficiency during experimental adjustments. 

Motion sickness in VR is primarily caused by discrepancies between physical sensations of movement and visual input lagging behind due to a low and/or unstable framerate \cite{effectOfFramerate2023}.
The susceptibility and severity of motion sickness is increased when the head is moved rapidly, such as during movement through the VR environment.
Contrary to this, during simulation interaction where the head remains in a relatively fixed position with the eyes focusing on specific elements, the impact of an unstable framerate is less pronounced.
Selective updating is beneficial in this context, as it conserves potentially scarce system resources especially when the user is most susceptible to motion sickness. 

Many routine user interactions do not require updates to beam parameters, such as adjusting optical elements not hit by any beams, navigating the virtual lab, reading experimental instructions, using the virtual laptop’s UI, or analyzing measurement results through manual inspection or interfaces. 
To quantify the impact of routine user interactions on performance, we analyzed the overall execution time of the simulation over the entire duration of a user's play session.
In this context, a speedrun was conducted by a physics expert who designed all of femtoPro's levels and objectives, which include building, adjusting, and analyzing the previously benchmarked experimental setups.
He was familiar with femtoPro's environment and controls, and memorized all objectives, reducing the time spent on reading instructions and other activities that do not impact the simulation's outcomes.
Therefore, we can assume that the expert's speedrun involves less idle time of the simulation compared to a typical play session of an average user.
This establishes a lower bound for the total number of idle frames, providing a baseline estimate of the efficiency gains from selective updating.
The expert finished all objectives of all levels within 24~min. 
In ca. 60,000 frames of 100,000 total frames, i.e., 60~\% of overall playtime, the simulation was in idle mode.
The simulation maintained a low CPU workload in the remaining frames: the total execution time of the simulation was 7~s out of 24~min, which is roughly 0.5~\% of the overall playtime.

Lower CPU resource usage, resulting from minimal simulation workload also reduces battery drain on mobile VR devices.
The effect becomes increasingly pronounced with the proportion of time the simulation operates under a reduced workload.
The impact of selective updating on power consumption was assessed using the Pico~3 mobile VR headset featuring a battery capacity of 19.61~Wh \cite{pico3Specs}.
With the simulation using all available resources each frame, power draw continuously reached 11.7~W.
Under the assumption that CPU resources are not internally throttled at lower battery levels to conserve power, this results in a total playtime of approximately 1~h 41~min from a fully charged to a completely drained battery state.
Conversely, with the simulation continuously in idle mode, the power draw was decreased by ca. 10~\%, resulting in 11 more minutes of playtime before the battery is depleted.
Consequently, by implementing the selective updating strategy, the frequency and length of recharging breaks is reduced, ensuring longer play sessions without interruption, and thus, offering a smoother experience for the user.

While aforementioned observations and results require validation through broader user studies, they offer a rough estimate of the benefits of selective updating on simulation efficiency in practice, complementing the experiment-specific benchmark results.

%
%
\section{Conclusion and future work}
\label{sec:conclusion}

We introduced the algorithmic requirements of femtoPro, a real-time interactive laser laboratory, capable of simulating pulsed or continuous-wave laser beams traversing through various optical elements.
Beams feature geometrical and wave-optics characteristics that are modified by optical elements through linear and second-order nonlinear optical responses.
We used a dynamic graph-based model to represent optical elements as nodes and beam segments between nodes as directed edges.
A major challenge in calculating this graph is the interdependent relationship between the segments of beam paths, which can create complex branched topologies. 
This dependency makes the simulation's outcomes highly sensitive to user interactions.

We implemented a custom-tailored path solver algorithm to meet femtoPro's requirements and associated challenges.
Every frame, the graph's topology is updated to reflect any changes introduced by user interactions. 
This is achieved through a node processing loop that successively generates each node's outgoing edges by applying linear and nonlinear optical transformations to its incident edges.
With femtoPro's freely configurable lab environment, beams can be aligned in various ways that might, in an extreme case, create infinite cycles of reflecting laser light, presenting a significant challenge to real-time calculations. 
To address this challenge, we implemented a cap on the number of optical responses that can be processed per frame.

A key feature of our algorithm are incremental updates, i.e., the prevention of unnecessary recalculations of parts of the graph that have not changed compared to last frame, which we refer to as ``selective updating''. 
Since many typical user interactions do not require rebuilding the entire graph, selective updating is a viable strategy for preserving computational resources.
We demonstrated that this approach reduces battery drain on mobile VR devices and helps maintaining a stable frame rate for a smooth and uninterrupted user experience.

We derived equations to estimate the simulation's performance costs, including memory consumption and CPU execution time.
This allows us to verify whether any particular experimental setup can be calculated in real time, thus streamlining the process of setting up predefined experiments.
We addressed excessive memory consumption by implementing a pooling system for edges and in-place operations for physics calculations.
Due to our efficient memory management and sufficiently large available RAM, memory capacity is not a concern in practice.
Consequently, our analysis is primarily focused on the simulation's execution time.
Based on developed equations we determined the runtime complexity, which, with respect to given input of edges and nodes, scales linearly with linear and quadratically with nonlinear optical responses.

We ran a detailed benchmark analysis of typical experimental setups, investigating the scalability and performance of the path solver algorithm in practice.
The results indicate that, although nonlinear optical responses tend to increase execution time, their impact on overall performance cost remains manageable.
Instead, beams that originate from lower parts of the graph and are reflected back towards the laser source are identified as performance bottleneck.
Such beam alignments require recalculations in unintended areas or even the entire graph, limiting the effectiveness of selective updating.

In a future version of femtoPro, we plan for the simulation to operate independently of frame time constraints.
If calculations take longer than one frame, we will present an intermediate result with reduced accuracy, ensuring the experience remains immersive and interactive.
This approach would be advantageous for other planned features with potentially high resource demands, such as the physically accurate treatment of cavities or the implementation of third-order nonlinear optical responses.

femtoPro successfully delivers real-time performance for current educational objectives in various freely configurable experimental setups, providing a valuable supplement to physics curricula.
We hope that the methodologies, solutions, and insights presented in this paper will allow researchers and developers in this realm to enhance their own simulation frameworks and foster innovation in the application of dynamic graph-based models in real-time interactive environments.

\section*{Statements and Declarations}
\subsection{Acknowledgements}
We acknowledge Tobias Buhl, Rahul Das, Andreas Knote, Maximilian Rall, Wilhelm Schnepp, Yannik Stamm, Samuel Truman, Anne Vetter, Theresa Weiglein, and Peter Ziegler for coding contributions.
\subsection{Funding}
This work was supported by the Fonds der Chemischen Industrie (FCI) and by the Julius-Maximilians-Universit{\"a}t W{\"u}rzburg Project WueDive via Stiftung Innovation in der Hochschullehre.
\subsection{Conflict of interest}
``femtoPro'' is a registered trademark of the University of W{\"u}rzburg. The software is available for free for both educational and commercial usage via the femtoPro website \cite{femtoProWebsite}. There are no commercial interests or partnerships established.
\subsection{Ethics approval and consent to participate}
Not applicable
\subsection{Consent for publication}
All participants involved in this paper provided consent for the publication of the findings and any potentially identifiable data included in this article.
\subsection{Data availability}
The raw data that support the findings of this paper are openly available in Zenodo at https://doi.org/10.5281/zenodo.15533468 \cite{zenodo2025}.
\subsection{Materials availability}
A demonstration trailer of the femtoPro app is available at the femtoPro website \cite{femtoProWebsite}.
\subsection{Code availability}
The latest femtoPro version 0.7.0 features  26 tutorial levels and is freely available for download on the femtoPro website \cite{femtoProWebsite}.
\subsection{Author contribution}
T.B. derived equations and coded the physical model.
A.M. wrote the paper, developed the path solver algorithm, coded the femtoPro core software and framework, and conducted the performance analysis.
S.M. coded didactic missions, participated in testing, and analyzed the validity of the physics simulation.
T.B. and S.v.M. supervised the project. All authors discussed the results and edited the manuscript.

\bibliographystyle{elsarticle-num} 
\bibliography{article}

\end{document}